\documentclass[aps,prl,twocolumn,superscriptaddress,nofootinbib,floatfix,noshowpacs]{revtex4-1}
\pdfoutput=1
\newif\ifcomment
\usepackage{amsmath,amssymb}
\usepackage{color,hyperref,url}
\usepackage{listings}
\usepackage{slashed}
\usepackage[pdftex]{graphicx}
\usepackage{epstopdf}
\usepackage{epsfig}
\usepackage{grffile}
\usepackage{relsize}
\graphicspath{{./img/}}
\usepackage{soul}
\newcommand{\beq}{\begin{equation}}
\newcommand{\eeq}{\end{equation}}
\newcommand{\ba}{\begin{array}}
\newcommand{\ea}{\end{array}}
\newcommand{\bea}{\begin{align}}
\newcommand{\eea}{\end{align}}
\newcommand{\bi}{\begin{itemize}}
\newcommand{\ei}{\end{itemize}}
\newcommand{\ben}{\begin{enumerate}}
\newcommand{\een}{\end{enumerate}}
\newcommand{\bc}{\begin{center}}
\newcommand{\ec}{\end{center}}
\newcommand{\bl}{\begin{flushleft}}
\newcommand{\el}{\end{flushleft}}
\newcommand{\br}{\begin{flushright}}
\newcommand{\er}{\end{flushright}}

\begin{document}
\title{Sieving parton distribution function moments via the moment problem}
\author{Xiaobin Wang} \email{wangxiaobin@mail.nankai.edu.cn}
\affiliation{School of Physics, Nankai University, Tianjin 300071, China}
\author{Minghui Ding} \email{m.ding@hzdr.de}
\affiliation{Helmholtz-Zentrum Dresden-Rossendorf, Bautzner Landstra{\ss}e 400, 01328 Dresden, Germany}
\author{Lei Chang} \email{leichang@nankai.edu.cn}
\affiliation{School of Physics, Nankai University, Tianjin 300071, China}
\date{\today}
\begin{abstract}
We apply a classical mathematical problem, the moment problem, with its related mathematical achievements, to the study of the parton distribution function (PDF) in hadron physics, and propose a strategy to sieve the moments of the PDF by exploiting its properties such as continuity, unimodality, and symmetry.
Through an error-inclusive sifting process, we refine three sets of PDF moments from Lattice QCD. This refinement significantly reduces the errors, particularly for higher order moments, and locates the peak of PDF simultaneously. As our strategy is universally applicable to PDF moments from any method, we strongly advocate its integration into all PDF moment calculations.
\end{abstract}
\maketitle

{\it Introduction.}---
Tossing a coin, it is well known that the probability of a head-up toss is the Bernoulli distribution, from which the mean, variance, skewness and kurtosis, i.e. the first to fourth standardised moments of the distribution can be obtained, which describe the different characteristics of the distribution, i.e., the average, the dispersion, the asymmetry, and the tailedness, respectively. Obtaining the moments from the distribution is straightforward, whereas the reverse is much more complex, i.e., whether or not a probability distribution can be uniquely determined by the sequence of moments, which is in fact a long-standing classical mathematical problem - the ``Moment Problem".


The one-dimensional (or univariate) moment problem has been well understood from several different perspectives (e.g., matrix theory, operator theory, probability theory and optimisation theory), after a great deal of innovative work by such eminent mathematicians as P.L.~Chebyshev (1874), A.A.~Markov (1884), T.J.~Stieltjes (1894), H.~Hamburger (1920), and F.~Hausdorff (1920) etc. Mathematicians have given answers to questions associated with the probability distribution obtained from moments, such as sufficient and necessary conditions for its existence, whether it is unique, and how to compute it.

This has greatly facilitated the application of the moment problem, and the scope of its illumination is extraordinary. In mathematics, the moment problem is closely related to the development of many branches of mathematics (functional analysis, function theory, real algebraic geometry, spectral theory, optimisation, numerical analysis, convex analysis, harmonic analysis, etc.)~\cite{landau1987moments,lasserre2009moments}. In economics, the moment problem has been applied to evaluate the risks brought by emergencies~\cite{tian2017moment,cornilly2018upper} and the price of a derivative security~\cite{lasserre2009moments}, the latter being a central question in financial economics, for which the economists Robert Merton and Myron Scholes were awarded the 1997 Nobel Prize. In engineering, the moment problem has been applied to deal with optimal control problems, admissible control problems, etc.~\cite{lasserre2009moments,rivero2010admissible,georgiou2017likelihood,karlsson2016multidimensional}. In physics, moments appear in many more places, especially in statistical physics, and whenever one encounters a probability distribution function that describes a practical phenomenon, it can be transformed into a treatment of its moments. 

In hadron physics, there is an important probability distribution function that reveals the internal structure of the hadron~\cite{Feynman:1969ej}, the parton distribution function (PDF), which describes the probability that a parton (quark and gluon) carries a certain fraction of the hadron's light-front momentum. Both historically and today, the study of the PDF occupies an extremely special place in hadron physics: for example, it led to the discovery of the quark, for which the experimental physicists involved were awarded the 1990 Nobel Prize, and it advanced the development of Quantum Chromodynamics (QCD). This means that any new understanding of the PDF is significant~\cite{Roberts:2021nhw}. Additionally, a range of high-energy, high-luminosity experimental facilities around the world are dedicated to the goal of accurately extracting hadron PDFs, including the Electron-Ion Collider (EIC)~\cite{Accardi:2012qut}, the Electron-ion collider in China (EicC)~\cite{Anderle:2021wcy}, and others.

%
%

Particularly, it is noteworthy that, as stated above, based on the nature of its probability distribution, the PDF is closely related to the mathematical moment problem, which is generally not well recognised at present~\cite{Mezrag:2023nkp}. In fact, for a long time,  theoretical calculations on the PDF have been limited to its lowest orders of moments, while its shape was unknown, i.e., historically, the knowledge of its moments was considerably ahead of knowledge of its shape~\cite{Best:1997qp}. Therefore, in this letter, we will study the moments of the PDF in terms of the mathematical moment problem. It is worth highlighting that, due to the fundamental and well-established nature of the moment problem, our study is highly general and rigorous. Specifically, our method is not limited to specific hadrons and partons, and thus is general and has a wide range of applications; our method is based on a solid mathematical foundation, and thus is highly rigorous.

{\it The moment problem and the sieve.}---
In this section, we will give some important conclusions about the moment problem, while their proofs are omitted. In particular, in order to facilitate the subsequent discussion of the PDF, the following mathematical contents are expressed using the probability density function instead of the measure that mathematicians are accustomed to using (in fact, this may raise some additional issues that we ignore here). All of these mathematical contents can be found in Refs.~\cite{schmüdgen2020lectures,schmudgen2017moment}, as can the proofs omitted here and more on the moment problem.


Considering the characteristics of PDF, we are concerned here with a special class of moment problem. We call it the truncated Hausdorff moment problem on the interval $[0,1]$, where ``truncated" and ``Hausdorff" represent finite sequence length and finite integral interval, respectively. Specifically, this type of moment problem can be expressed as follows: given a sequence of real numbers $(s_k)_{k=0}^m$, it is required to find a non-negative function $f(x)$ that satisfies the equation:
\begin{equation}
    s_k=\int_0^1x^kf(x)\,dx\,,
\end{equation}
for $k=0,1,2,\dots,m$, where $f(x)$ is the PDF and $s_k$ is the moment of PDF. 
This problem is not always solvable; it requires that the given sequence satisfy some conditions. Before stating these conditions, we need to introduce some terminology and notations. If a solution to this problem exists, then the given sequence is called a truncated $[0,1]$-moment sequence. $A\succ0$ and $A\succeq0$ indicate that the matrix $A$ is positive definite and positive semidefinite, respectively. If the matrix $A$ is positive semidefinite but not positive definite, we say that the ``$\succeq$" in $A\succeq0$ takes the equal sign. $(s_{i+j})_{i,j=0}^n$ represents the following Hankel matrix:
\begin{equation}
    (s_{i+j})_{i,j=0}^n=\begin{pmatrix}
        s_0&s_1&\cdots&s_{n-1}&s_n\\
        s_1&s_2&\cdots&s_n&s_{n+1}\\
        \vdots&\vdots&\ddots&\vdots&\vdots\\
        s_{n-1}&s_n&\cdots&s_{2n-2}&s_{2n-1}\\
        s_n&s_{n+1}&\cdots&s_{2n-1}&s_{2n}\\
    \end{pmatrix}\,.
\end{equation}

Now we can show the important conditions for the existence of a solution.
The necessary and sufficient conditions for $(s_k)_{k=0}^m$ to be a truncated $[0,1]$-moment sequence are: for the even case $m=2n$,
\begin{equation} \label{nn1}
    (s_{i+j})_{i,j=0}^n\succeq0\ \quad \text{and}\ \quad (s_{i+j+1}-s_{i+j+2})_{i,j=0}^{n-1}\succeq0\,;
\end{equation}
for the odd case $m=2n+1$,
\begin{equation} \label{nn2}
    (s_{i+j+1})_{i,j=0}^n\succeq0\ \quad \text{and}\ \quad (s_{i+j}-s_{i+j+1})_{i,j=0}^n\succeq0\,.
\end{equation}
In addition, if any ``$\succeq$" in the above conditions takes the equal sign, then the corresponding $f(x)$ is unique and consists of a series of delta functions.

These conditions are not trivial; in fact, they are difficult to satisfy when $m$ is large. Using these constraints, we can eliminate from moments the data that do not satisfy these conditions and retain the data that do. In this process, these conditions act like a sieve, so in the following we will use the term ``sieve" to refer to these conditions. More importantly, when applied to PDF moments, this sieve can be further strengthened due to the special properties of PDF itself.

{\it Strengthening of the sieve.}---
In the above discussion, we have only considered the non-negativity of PDF. However  in fact, the PDF usually has more useful properties. Next, we will consider three possible properties of the PDF and apply them to give constraints on its moments, thus strengthening the sieve accordingly. It should be noted in advance that the discussions in this part are progressive, i.e., the cases discussed later always have all of their aforementioned properties at the same time. Since the proofs covered in this part do not appear in the two mathematical references mentioned above,
%
%
the proofs of necessity are given below, and the proofs of sufficiency are included in Appendix A.

The first property is continuity. Since the delta function is not continuous and the PDF generally does not contain delta functions at current scales, the Hankel matrix is not positive semidefinite but positive definite, i.e., the sieve after adding continuity is Eqs.~\eqref{nn1}, \eqref{nn2}, with all constraints being ``$\succ0$".


The second property is unimodality. It specifically requires $f(x)$ to satisfy the following conditions:
\begin{subequations}
\begin{align}
&f(0)=f(1)=0\,,\label{ff}\\
    &\exists\lambda\in(0,1),\quad f'(x)
    \begin{cases}
        \ge0,&\,\,0<x<\lambda\\
        \le0,&\,\,\lambda<x<1\\
    \end{cases}\,.\label{el}
\end{align}
\end{subequations}
Unlike continuity, only certain PDFs satisfy the above conditions. The $f(x)$ that satisfies the above condition is unimodal, with the peak location being $\lambda$. We therefore refer to this property as ``unimodality".
Using Eq.~\eqref{ff} and integration by parts, we get:
\begin{equation}
    t_k=\int_0^1x^kf'(x)\,dx=-ks_{k-1}\,,
\end{equation}
for $k=0,1,2,\dots,m+1$. 
%
%
Eq.~\eqref{el} indicates $g(x)=(\lambda-x)f'(x)\ge0$, with $g(x)$ being a continuous function, and the moments of $g(x)$ must likewise satisfy the constrains that the moments of $f(x)$ obey.
Subsequently, the sieve considering unimodality is: for the even case $m=2n$,
\begin{equation}\label{el1}
    \begin{aligned}
        &\exists\lambda\in(0,1)\,,\quad (\lambda t_{i+j}-t_{i+j+1})_{i,j=0}^n\succ0\\
        &\text{and}\ \quad \left[\lambda t_{i+j+1}-(\lambda+1)t_{i+j+2}+t_{i+j+3}\right]_{i,j=0}^{n-1}\succ0\,;
    \end{aligned}
\end{equation}
for the odd case $m=2n+1$,
\begin{equation}\label{el2}
    \begin{aligned}
        &\exists\lambda\in(0,1)\,,\quad (\lambda t_{i+j+1}-t_{i+j+2})_{i,j=0}^n\succ0\\
        &\text{and}\ \quad \left[\lambda t_{i+j}-(\lambda+1)t_{i+j+1}+t_{i+j+2}\right]_{i,j=0}^{n}\succ0\,.
    \end{aligned}
\end{equation}
In addition, in some specific cases, the peak location $\lambda$ is subject to some physically relevant constraints, and then the interval of $\lambda$ in the above two equations will need to be changed from $(0,1)$ to the corresponding constrained interval.


The third property is the symmetry. It means $f(x)=f(1-x)$ in some special cases, thus the peak location $\lambda$ is $1/2$. 
Consequently, the sieve after adding symmetry contains not only the aforementioned constraints, i.e., Eqs.~\eqref{el1},~\eqref{el2} with $\lambda=1/2$, but also an additional constraint on top of that,
\begin{equation}\label{kl}
    \sum_{k=0}^{2l+1}(-2)^{k}\binom{2l+1}{k}s_k=0\,,
\end{equation}
for $l=0,1,2,\dots,\left\lfloor\frac{m-1}{2}\right\rfloor$, where $\lfloor x\rfloor$ is the floor function.


{\it Application of the sieve.}---
Generally, the PDF moments obtained by some method of calculation are not completely accurate and may contain errors (except for the zero-order moment determined by normalization), the values of which take the form of, for example, $\mu_k\pm\sigma_k$, where $\mu_k$ is the mean value and $\sigma_k$ is the error. In this case, we cannot directly use the sieve to sift, but need to employ a sifting procedure that takes into account the errors. The procedure can be divided into three steps.

The first step is to generate the raw dataset of moments. We treat each moment as a random variable and assume that it obeys the standard normal distribution $N(\mu_k,\sigma_k^2)$, from which we randomly generate data for each moment. By putting a finite number of moments of interest together, we generate a dataset of moments. Particularly, when generating the raw dataset of moments corresponding to the symmetric PDF, it is important to note that the probability distributions to which the moments are subjected are not independent standard normal distributions, but rather the normal distributions with altered parameters, as detailed in Appendix B.

The second step is to sift the raw dataset of moments. An appropriate sieve must be selected based on the properties of the target PDF. For example, if the target PDF has non-negativity, then Eqs.~\eqref{nn1}, \eqref{nn2} can be used as the sieve; on top of that, add continuity, then all constraints are positive definite; add unimodality, then Eqs.~\eqref{el1}, \eqref{el2} is the sieve; and add symmetry, then Eq.~\eqref{kl} and Eqs.~\eqref{el1}, \eqref{el2} with $\lambda=1/2$ is the sieve.  With a suitable sieve, if the raw dataset generated in the first step meets the sieving criteria, it will be retained as a sample; otherwise, it will be discarded.

The third step is to accumulate the samples and then calculate the overall mean and error for each moment. The first two steps are repeated until the number of samples reaches the pre-set sample size, and then the overall mean and error are calculated for each moment in the sample set. Particularly, for the unimodal case, we can also calculate the peak location for each sample, and count all the samples to give the possible range of the unimodal peak, as well as the overall mean and error of the lower and upper bounds of that range.

{\it Results.}---
In fact, the above sieve and sifting procedure are very general, i.e., they can be used to sift the PDF moments of any hadron computed by any method. Here, we illustrate the practical application of our proposed sieve and sifting procedure by taking the PDF moments computed by Lattice QCD from the first principle as an example. We have selected three sets of Lattice QCD moments of the light quark valence PDF calculated at $2$ GeV, as shown in Table~\ref{na}. Among them, $\mathcal{A}$ and $\mathcal{B}$ are the moments of the pion PDF, with data from Ref.~\cite{PhysRevD.100.114512} and Ref.~\cite{Alexandrou:2021ejy}, respectively; and $\mathcal{C}$ is the moments of the kaon PDF, with data from Ref.~\cite{Alexandrou:2021ejy}.

The PDF at the hadron scale $\zeta_{\mathcal{H}}$ - a specific scale at which valence quarks/antiquarks in the hadron carry all hadron momentum - has more properties~\cite{Cui:2022bxn}, e.g., the pion PDF has continuity, unimodality and symmetry. The kaon PDF has continuity and unimodality, and the peak position $\lambda$ should be less than $1/2$ because the mass of the $u$ quark is less than the mass of the $s$ quark. Thus, at this specific hadron scale, one can use the strengthened version of the sieve resulting from the PDF properties, to better reflect the practical use of our proposed sieve. Therefore, in order to maximize the use of the sieve, we use the all-orders evolution method~\cite{Raya:2021zrz,Yin:2023dbw} to evolve the Lattice QCD moments listed in Table~\ref{na} to 
the hadron scale. 
Likewise, the error in the moment evolves with the scale; we compute the error according to the standard error propagation formula. The methods for all-orders evolution and error evolution are detailed in Appendix C.

\begin{table}[t]
\caption{Lattice QCD moments of the light quark valence PDF calculated at $2$ GeV, $\mathcal{A}$ and $\mathcal{B}$ are the moments of the pion PDF, and $\mathcal{C}$ is the moments of the kaon PDF.}\label{na}
\begin{ruledtabular}
    \begin{tabular}{cccc}
        $m$&$\mathcal{A}$ \cite{PhysRevD.100.114512}&$\mathcal{B}$ \cite{Alexandrou:2021ejy}&$\mathcal{C}$ \cite{Alexandrou:2021ejy}\\
        \hline
        1&0.2541(26)&0.261(3)&0.246(2)\\
        2&0.094(12)&0.110(7)&0.096(2)\\
        3&0.057(4)&0.024(18)&0.033(6)\\
        4&0.015(12)&-&-\\
    \end{tabular}
\end{ruledtabular}
\end{table}

\begin{figure}[t]
    \centering
    \includegraphics[width=\linewidth]{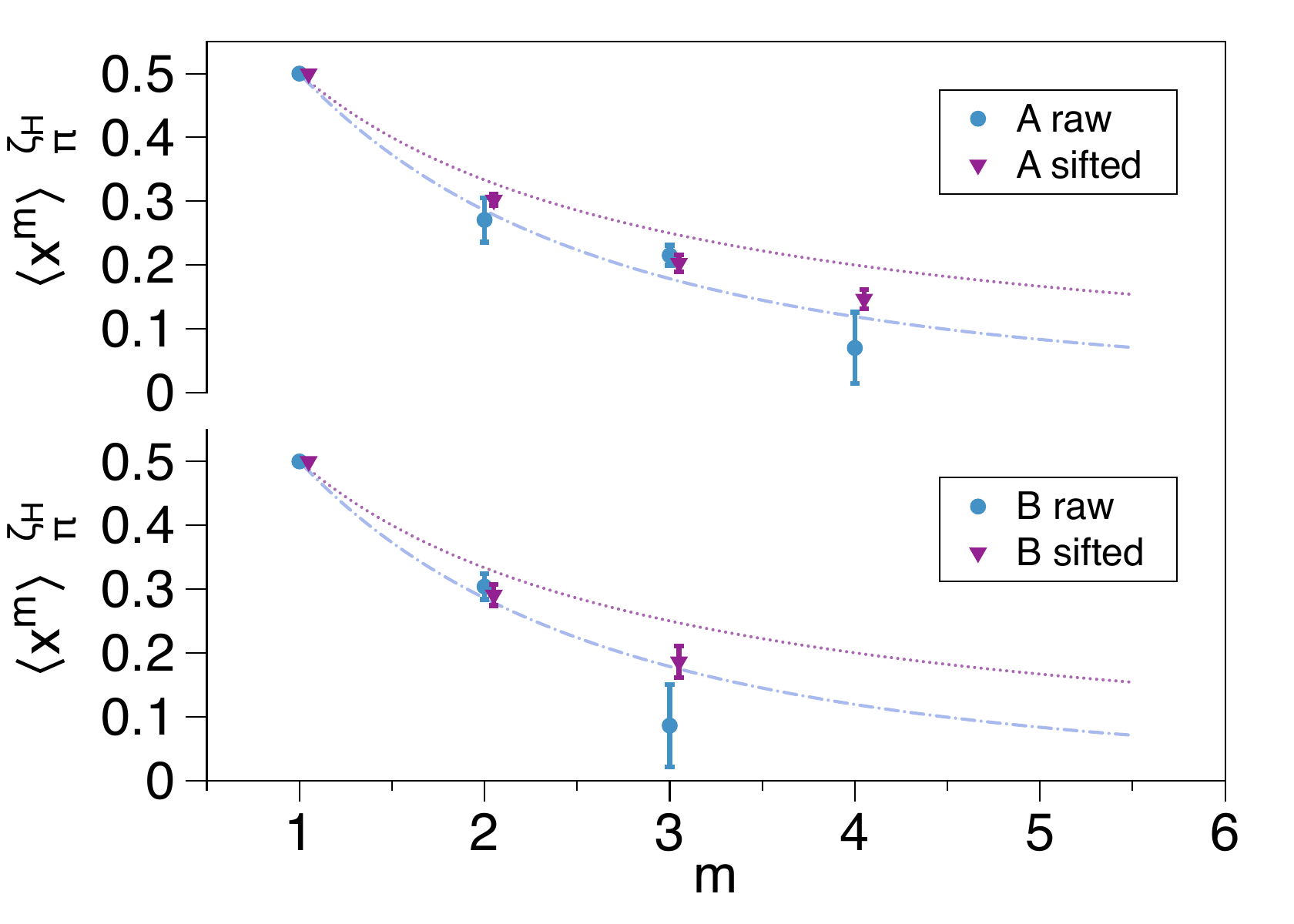}
    \caption{The sifting of pion PDF moments at hadron scale $\zeta_{\mathcal{H}}$.
    The azure dots and error bars are raw data, and the plum inverted triangles and error bars are sifted data. The azure dotted dashed curve is the moment of $30x^2(1-x)^2$, and the plum dotted curve is the moment of $1$.}
    \label{ab}
\end{figure}

\begin{figure}[t]
    \centering
    \includegraphics[width=\linewidth]{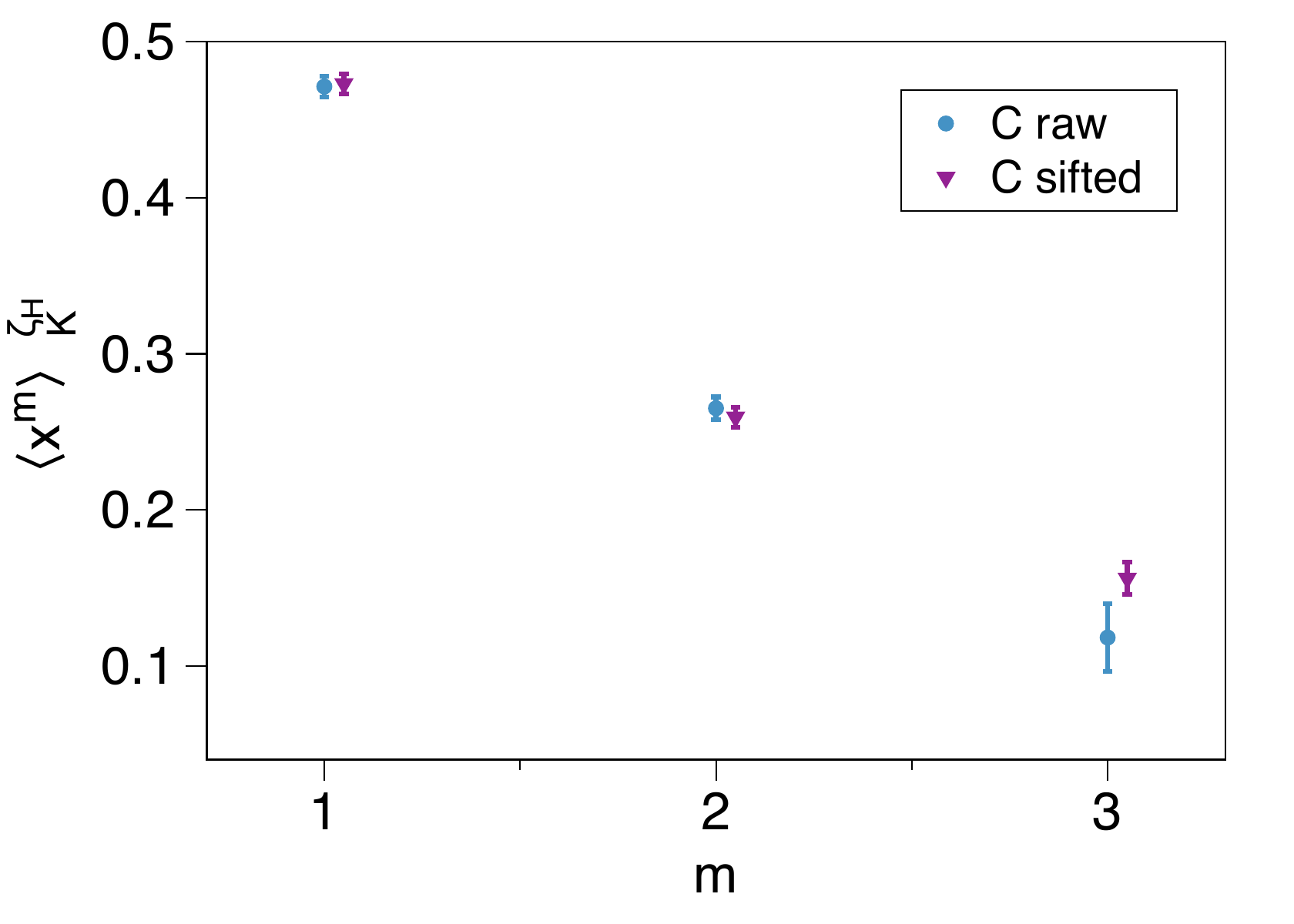}
    \caption{The sifting of kaon PDF moments at hadron scale $\zeta_{\mathcal{H}}$.
    The azure dots and error bars are raw data, and the plum inverted triangles and error bars are sifted data.}
    \label{c}
\end{figure}

%
%
The raw data are listed in Figs.~\ref{ab}, \ref{c} and labelled as A raw, B raw and C raw, respectively. We then sift the data according to the sifting procedure proposed above, setting the sample size to $1000$, and after sifting, we obtain the sifted data labelled as A sifted, B sifted and C sifted, respectively. To avoid overlapping, the horizontal coordinates of the sifted data are shifted to the right by $0.05$ units, i.e., $m+0.05$.

We find a significant reduction in the errors of the sifted data as detailed in Table.~\ref{ma}. Isospin symmetry ensures that the first moment of the pion PDF at hadron scale is $1/2$. Our sieve effectively reduces the errors for all moments except it, and the effect is even more pronounced at higher orders, e.g., its reductions in errors for the second and fourth moments in $\mathcal{A}$, the third moment in $\mathcal{B}$, and the third moment in $\mathcal{C}$ are significant at $74.2\%$, $73.3\%$, $61.6\%$ and $52.0\%$, respectively.

Additionally, the sifted data not only have reduced errors, but their values are more in line with physics. Specifically, there are two apparent limits for the Goldstone boson PDF at the hadron scale, one in the scale-free form, $30x^2(1-x)^2$, and the other as the result of the massless pion using contact interactions, $1$. These two limits give rise to the moment boundaries as shown in Fig.~\ref{ab}. Notably, the mean of the fourth moment in $\mathcal{A}$ and the mean of the third moment in $\mathcal{B}$, whose raw data are clearly not in the region between the two boundaries, enter the region after being sifted.

\begin{table}[t]
\caption{Percentage reduction in errors after sifting.}\label{ma}
\begin{ruledtabular}
    \begin{tabular}{cccc}
        $m$&$\mathcal{A}$&$\mathcal{B}$&$\mathcal{C}$\\
        \hline
        1&-&-&2.2\%\\
        2&74.2\%&17.8\%&12.8\%\\
        3&14.3\%&61.6\%&52.0\%\\
        4&73.3\%&-&-\\
    \end{tabular}
\end{ruledtabular}
\end{table}

Moreover, the sieve provides the ability to accurately determine the peak location of the PDF. For example, considering kaon PDF at the hadron scale, it is worth noting that $\lambda$ should be limited to less than $1/2$. We obtain the lower and upper limits of $\lambda$, which are $0.27(11)$ and $0.498(13)$, respectively. Due to the limited number of moments, there is still a large uncertainty, and considering more moments can lead to more accurate bounds.

{\it Summary.}---
We study moments of the PDF in the context of the moment problem. Our study finds that owing to the non-negativity of the PDF, moments are subject to a collection of intricate constraints, which are referred to as the ``sieve". Furthermore, the effectiveness of the sieve is enhanced in cases where the PDF exhibits features such as continuity, unimodality, and symmetry. To address this, we design an error-inclusive, random process-based sifting procedure.

The proposed sieve and sifting procedure are general and applicable to PDF moments of any hadron, regardless of the method used or the scale at which they are obtained. As a practical application, we select three sets of Lattice-QCD results for sifting. Since the PDF at the hadron scale may have more properties, we evolve the Lattice-QCD moments at $2$ GeV to the hadron scale. After the sifting, for pion, the errors of all moments are significantly reduced, especially for higher-order moments, and the mean values of the moments are more in line with physics; for kaon, the range of the peak location of its PDF is obtained. These results demonstrate the power of our proposed sieve and sifting procedure, and we therefore strongly recommend the integration of this powerful and general sieve and sifting procedure into all calculations of PDF moments. Furthermore, this conclusion extends to the study of, for example, particle distribution amplitudes, and other more generalized distribution functions.

{\it Acknowledgments.}---
Work supported by National Natural Science Foundation of China (grant no. 12135007). MD is grateful for support by Helmholtz-Zentrum Dresden-Rossendorf High Potential Programme.

\bibliography{mpNKU}

\begin{thebibliography}{20}%
\makeatletter
\providecommand \@ifxundefined [1]{%
 \@ifx{#1\undefined}
}%
\providecommand \@ifnum [1]{%
 \ifnum #1\expandafter \@firstoftwo
 \else \expandafter \@secondoftwo
 \fi
}%
\providecommand \@ifx [1]{%
 \ifx #1\expandafter \@firstoftwo
 \else \expandafter \@secondoftwo
 \fi
}%
\providecommand \natexlab [1]{#1}%
\providecommand \enquote  [1]{``#1''}%
\providecommand \bibnamefont  [1]{#1}%
\providecommand \bibfnamefont [1]{#1}%
\providecommand \citenamefont [1]{#1}%
\providecommand \href@noop [0]{\@secondoftwo}%
\providecommand \href [0]{\begingroup \@sanitize@url \@href}%
\providecommand \@href[1]{\@@startlink{#1}\@@href}%
\providecommand \@@href[1]{\endgroup#1\@@endlink}%
\providecommand \@sanitize@url [0]{\catcode `\\12\catcode `\$12\catcode `\&12\catcode `\#12\catcode `\^12\catcode `\_12\catcode `\%12\relax}%
\providecommand \@@startlink[1]{}%
\providecommand \@@endlink[0]{}%
\providecommand \url  [0]{\begingroup\@sanitize@url \@url }%
\providecommand \@url [1]{\endgroup\@href {#1}{\urlprefix }}%
\providecommand \urlprefix  [0]{URL }%
\providecommand \Eprint [0]{\href }%
\providecommand \doibase [0]{http://dx.doi.org/}%
\providecommand \selectlanguage [0]{\@gobble}%
\providecommand \bibinfo  [0]{\@secondoftwo}%
\providecommand \bibfield  [0]{\@secondoftwo}%
\providecommand \translation [1]{[#1]}%
\providecommand \BibitemOpen [0]{}%
\providecommand \bibitemStop [0]{}%
\providecommand \bibitemNoStop [0]{.\EOS\space}%
\providecommand \EOS [0]{\spacefactor3000\relax}%
\providecommand \BibitemShut  [1]{\csname bibitem#1\endcsname}%
\let\auto@bib@innerbib\@empty
\bibitem [{\citenamefont {Landau}(1987)}]{landau1987moments}%
  \BibitemOpen
  \bibfield  {author} {\bibinfo {author} {\bibfnamefont {H.~J.}\ \bibnamefont {Landau}},\ }\href@noop {} {\emph {\bibinfo {title} {Moments in mathematics}}},\ Vol.~\bibinfo {volume} {37}\ (\bibinfo  {publisher} {American Mathematical Soc.},\ \bibinfo {year} {1987})\BibitemShut {NoStop}%
\bibitem [{\citenamefont {Lasserre}(2009)}]{lasserre2009moments}%
  \BibitemOpen
  \bibfield  {author} {\bibinfo {author} {\bibfnamefont {J.~B.}\ \bibnamefont {Lasserre}},\ }\href@noop {} {\emph {\bibinfo {title} {Moments, positive polynomials and their applications}}},\ Vol.~\bibinfo {volume} {1}\ (\bibinfo  {publisher} {World Scientific},\ \bibinfo {year} {2009})\BibitemShut {NoStop}%
\bibitem [{\citenamefont {Tian}\ \emph {et~al.}(2017)\citenamefont {Tian}, \citenamefont {Cox},\ and\ \citenamefont {Zuluaga}}]{tian2017moment}%
  \BibitemOpen
  \bibfield  {author} {\bibinfo {author} {\bibfnamefont {R.}~\bibnamefont {Tian}}, \bibinfo {author} {\bibfnamefont {S.}~\bibnamefont {Cox}}, \ and\ \bibinfo {author} {\bibfnamefont {L.}~\bibnamefont {Zuluaga}},\ }\href {\doibase 10.1080/10920277.2017.1302805} {\bibfield  {journal} {\bibinfo  {journal} {North American Actuarial Journal}\ }\textbf {\bibinfo {volume} {21}},\ \bibinfo {pages} {1} (\bibinfo {year} {2017})}\BibitemShut {NoStop}%
\bibitem [{\citenamefont {Cornilly}\ \emph {et~al.}(2018)\citenamefont {Cornilly}, \citenamefont {Rueschendorf},\ and\ \citenamefont {Vanduffel}}]{cornilly2018upper}%
  \BibitemOpen
  \bibfield  {author} {\bibinfo {author} {\bibfnamefont {D.}~\bibnamefont {Cornilly}}, \bibinfo {author} {\bibfnamefont {L.}~\bibnamefont {Rueschendorf}}, \ and\ \bibinfo {author} {\bibfnamefont {S.}~\bibnamefont {Vanduffel}},\ }\href {\doibase 10.1016/j.insmatheco.2018.07.002} {\bibfield  {journal} {\bibinfo  {journal} {Insurance: Mathematics and Economics}\ }\textbf {\bibinfo {volume} {82}},\ \bibinfo {pages} {141} (\bibinfo {year} {2018})}\BibitemShut {NoStop}%
\bibitem [{\citenamefont {Rivero}\ \emph {et~al.}(2010)\citenamefont {Rivero}, \citenamefont {Korobov},\ and\ \citenamefont {Sklyar}}]{rivero2010admissible}%
  \BibitemOpen
  \bibfield  {author} {\bibinfo {author} {\bibfnamefont {A.}~\bibnamefont {Rivero}}, \bibinfo {author} {\bibfnamefont {V.}~\bibnamefont {Korobov}}, \ and\ \bibinfo {author} {\bibfnamefont {G.}~\bibnamefont {Sklyar}},\ }\href {\doibase 10.1016/j.aml.2009.06.030} {\bibfield  {journal} {\bibinfo  {journal} {Appl. Math. Lett.}\ }\textbf {\bibinfo {volume} {23}},\ \bibinfo {pages} {58} (\bibinfo {year} {2010})}\BibitemShut {NoStop}%
\bibitem [{\citenamefont {Georgiou}\ and\ \citenamefont {Lindquist}(2016)}]{georgiou2017likelihood}%
  \BibitemOpen
  \bibfield  {author} {\bibinfo {author} {\bibfnamefont {T.}~\bibnamefont {Georgiou}}\ and\ \bibinfo {author} {\bibfnamefont {A.}~\bibnamefont {Lindquist}},\ }\href {\doibase 10.1109/TAC.2017.2672862} {\bibfield  {journal} {\bibinfo  {journal} {IEEE Transactions on Automatic Control}\ }\textbf {\bibinfo {volume} {PP}} (\bibinfo {year} {2016}),\ 10.1109/TAC.2017.2672862}\BibitemShut {NoStop}%
\bibitem [{\citenamefont {Karlsson}\ \emph {et~al.}(2015)\citenamefont {Karlsson}, \citenamefont {Lindquist},\ and\ \citenamefont {Ringh}}]{karlsson2016multidimensional}%
  \BibitemOpen
  \bibfield  {author} {\bibinfo {author} {\bibfnamefont {J.}~\bibnamefont {Karlsson}}, \bibinfo {author} {\bibfnamefont {A.}~\bibnamefont {Lindquist}}, \ and\ \bibinfo {author} {\bibfnamefont {A.}~\bibnamefont {Ringh}},\ }\href {\doibase 10.1007/s00020-015-2248-z} {\bibfield  {journal} {\bibinfo  {journal} {Integral Equations and Operator Theory}\ }\textbf {\bibinfo {volume} {84}} (\bibinfo {year} {2015}),\ 10.1007/s00020-015-2248-z}\BibitemShut {NoStop}%
\bibitem [{\citenamefont {Feynman}(1969)}]{Feynman:1969ej}%
  \BibitemOpen
  \bibfield  {author} {\bibinfo {author} {\bibfnamefont {R.~P.}\ \bibnamefont {Feynman}},\ }\href {\doibase 10.1103/PhysRevLett.23.1415} {\bibfield  {journal} {\bibinfo  {journal} {Phys. Rev. Lett.}\ }\textbf {\bibinfo {volume} {23}},\ \bibinfo {pages} {1415} (\bibinfo {year} {1969})}\BibitemShut {NoStop}%
\bibitem [{\citenamefont {Roberts}\ \emph {et~al.}(2021)\citenamefont {Roberts}, \citenamefont {Richards}, \citenamefont {Horn},\ and\ \citenamefont {Chang}}]{Roberts:2021nhw}%
  \BibitemOpen
  \bibfield  {author} {\bibinfo {author} {\bibfnamefont {C.~D.}\ \bibnamefont {Roberts}}, \bibinfo {author} {\bibfnamefont {D.~G.}\ \bibnamefont {Richards}}, \bibinfo {author} {\bibfnamefont {T.}~\bibnamefont {Horn}}, \ and\ \bibinfo {author} {\bibfnamefont {L.}~\bibnamefont {Chang}},\ }\href {\doibase 10.1016/j.ppnp.2021.103883} {\bibfield  {journal} {\bibinfo  {journal} {Prog. Part. Nucl. Phys.}\ }\textbf {\bibinfo {volume} {120}},\ \bibinfo {pages} {103883} (\bibinfo {year} {2021})},\ \Eprint {http://arxiv.org/abs/2102.01765} {arXiv:2102.01765 [hep-ph]} \BibitemShut {NoStop}%
\bibitem [{\citenamefont {Accardi}\ \emph {et~al.}(2016)\citenamefont {Accardi} \emph {et~al.}}]{Accardi:2012qut}%
  \BibitemOpen
  \bibfield  {author} {\bibinfo {author} {\bibfnamefont {A.}~\bibnamefont {Accardi}} \emph {et~al.},\ }\href {\doibase 10.1140/epja/i2016-16268-9} {\bibfield  {journal} {\bibinfo  {journal} {Eur. Phys. J. A}\ }\textbf {\bibinfo {volume} {52}},\ \bibinfo {pages} {268} (\bibinfo {year} {2016})},\ \Eprint {http://arxiv.org/abs/1212.1701} {arXiv:1212.1701 [nucl-ex]} \BibitemShut {NoStop}%
\bibitem [{\citenamefont {Anderle}\ \emph {et~al.}(2021)\citenamefont {Anderle} \emph {et~al.}}]{Anderle:2021wcy}%
  \BibitemOpen
  \bibfield  {author} {\bibinfo {author} {\bibfnamefont {D.~P.}\ \bibnamefont {Anderle}} \emph {et~al.},\ }\href {\doibase 10.1007/s11467-021-1062-0} {\bibfield  {journal} {\bibinfo  {journal} {Front. Phys. (Beijing)}\ }\textbf {\bibinfo {volume} {16}},\ \bibinfo {pages} {64701} (\bibinfo {year} {2021})},\ \Eprint {http://arxiv.org/abs/2102.09222} {arXiv:2102.09222 [nucl-ex]} \BibitemShut {NoStop}%
\bibitem [{\citenamefont {Mezrag}(2023)}]{Mezrag:2023nkp}%
  \BibitemOpen
  \bibfield  {author} {\bibinfo {author} {\bibfnamefont {C.}~\bibnamefont {Mezrag}},\ }\href {\doibase 10.3390/particles6010015} {\bibfield  {journal} {\bibinfo  {journal} {Particles}\ }\textbf {\bibinfo {volume} {6}},\ \bibinfo {pages} {262} (\bibinfo {year} {2023})}\BibitemShut {NoStop}%
\bibitem [{\citenamefont {Best}\ \emph {et~al.}(1997)\citenamefont {Best}, \citenamefont {Gockeler}, \citenamefont {Horsley}, \citenamefont {Ilgenfritz}, \citenamefont {Perlt}, \citenamefont {Rakow}, \citenamefont {Schafer}, \citenamefont {Schierholz}, \citenamefont {Schiller},\ and\ \citenamefont {Schramm}}]{Best:1997qp}%
  \BibitemOpen
  \bibfield  {author} {\bibinfo {author} {\bibfnamefont {C.}~\bibnamefont {Best}}, \bibinfo {author} {\bibfnamefont {M.}~\bibnamefont {Gockeler}}, \bibinfo {author} {\bibfnamefont {R.}~\bibnamefont {Horsley}}, \bibinfo {author} {\bibfnamefont {E.-M.}\ \bibnamefont {Ilgenfritz}}, \bibinfo {author} {\bibfnamefont {H.}~\bibnamefont {Perlt}}, \bibinfo {author} {\bibfnamefont {P.~E.~L.}\ \bibnamefont {Rakow}}, \bibinfo {author} {\bibfnamefont {A.}~\bibnamefont {Schafer}}, \bibinfo {author} {\bibfnamefont {G.}~\bibnamefont {Schierholz}}, \bibinfo {author} {\bibfnamefont {A.}~\bibnamefont {Schiller}}, \ and\ \bibinfo {author} {\bibfnamefont {S.}~\bibnamefont {Schramm}},\ }\href {\doibase 10.1103/PhysRevD.56.2743} {\bibfield  {journal} {\bibinfo  {journal} {Phys. Rev. D}\ }\textbf {\bibinfo {volume} {56}},\ \bibinfo {pages} {2743} (\bibinfo {year} {1997})},\ \Eprint {http://arxiv.org/abs/hep-lat/9703014} {arXiv:hep-lat/9703014} \BibitemShut {NoStop}%
\bibitem [{\citenamefont {Schmüdgen}(2020)}]{schmüdgen2020lectures}%
  \BibitemOpen
  \bibfield  {author} {\bibinfo {author} {\bibfnamefont {K.}~\bibnamefont {Schmüdgen}},\ }\href@noop {} {\enquote {\bibinfo {title} {Ten lectures on the moment problem},}\ } (\bibinfo {year} {2020}),\ \Eprint {http://arxiv.org/abs/2008.12698} {arXiv:2008.12698 [math.FA]} \BibitemShut {NoStop}%
\bibitem [{\citenamefont {Schmüdgen}(2017)}]{schmudgen2017moment}%
  \BibitemOpen
  \bibfield  {author} {\bibinfo {author} {\bibfnamefont {K.}~\bibnamefont {Schmüdgen}},\ }\href@noop {} {\emph {\bibinfo {title} {The Moment Problem}}}\ (\bibinfo  {publisher} {Springer},\ \bibinfo {year} {2017})\BibitemShut {NoStop}%
\bibitem [{\citenamefont {Jo\'o}\ \emph {et~al.}(2019)\citenamefont {Jo\'o}, \citenamefont {Karpie}, \citenamefont {Orginos}, \citenamefont {Radyushkin}, \citenamefont {Richards}, \citenamefont {Sufian},\ and\ \citenamefont {Zafeiropoulos}}]{PhysRevD.100.114512}%
  \BibitemOpen
  \bibfield  {author} {\bibinfo {author} {\bibfnamefont {B.}~\bibnamefont {Jo\'o}}, \bibinfo {author} {\bibfnamefont {J.}~\bibnamefont {Karpie}}, \bibinfo {author} {\bibfnamefont {K.}~\bibnamefont {Orginos}}, \bibinfo {author} {\bibfnamefont {A.~V.}\ \bibnamefont {Radyushkin}}, \bibinfo {author} {\bibfnamefont {D.~G.}\ \bibnamefont {Richards}}, \bibinfo {author} {\bibfnamefont {R.~S.}\ \bibnamefont {Sufian}}, \ and\ \bibinfo {author} {\bibfnamefont {S.}~\bibnamefont {Zafeiropoulos}},\ }\href {\doibase 10.1103/PhysRevD.100.114512} {\bibfield  {journal} {\bibinfo  {journal} {Phys. Rev. D}\ }\textbf {\bibinfo {volume} {100}},\ \bibinfo {pages} {114512} (\bibinfo {year} {2019})}\BibitemShut {NoStop}%
\bibitem [{\citenamefont {Alexandrou}\ \emph {et~al.}(2022)\citenamefont {Alexandrou}, \citenamefont {Bacchio}, \citenamefont {Clo\"et}, \citenamefont {Constantinou}, \citenamefont {Hadjiyiannakou}, \citenamefont {Koutsou},\ and\ \citenamefont {Lauer}}]{Alexandrou:2021ejy}%
  \BibitemOpen
  \bibfield  {author} {\bibinfo {author} {\bibfnamefont {C.}~\bibnamefont {Alexandrou}}, \bibinfo {author} {\bibfnamefont {S.}~\bibnamefont {Bacchio}}, \bibinfo {author} {\bibfnamefont {I.}~\bibnamefont {Clo\"et}}, \bibinfo {author} {\bibfnamefont {M.}~\bibnamefont {Constantinou}}, \bibinfo {author} {\bibfnamefont {K.}~\bibnamefont {Hadjiyiannakou}}, \bibinfo {author} {\bibfnamefont {G.}~\bibnamefont {Koutsou}}, \ and\ \bibinfo {author} {\bibfnamefont {C.}~\bibnamefont {Lauer}},\ }\href {\doibase 10.22323/1.396.0169} {\bibfield  {journal} {\bibinfo  {journal} {PoS}\ }\textbf {\bibinfo {volume} {LATTICE2021}},\ \bibinfo {pages} {169} (\bibinfo {year} {2022})},\ \Eprint {http://arxiv.org/abs/2112.03952} {arXiv:2112.03952 [hep-lat]} \BibitemShut {NoStop}%
\bibitem [{\citenamefont {Cui}\ \emph {et~al.}(2022)\citenamefont {Cui}, \citenamefont {Ding}, \citenamefont {Morgado}, \citenamefont {Raya}, \citenamefont {Binosi}, \citenamefont {Chang}, \citenamefont {De~Soto}, \citenamefont {Roberts}, \citenamefont {Rodr\'\i{}guez-Quintero},\ and\ \citenamefont {Schmidt}}]{Cui:2022bxn}%
  \BibitemOpen
  \bibfield  {author} {\bibinfo {author} {\bibfnamefont {Z.~F.}\ \bibnamefont {Cui}}, \bibinfo {author} {\bibfnamefont {M.}~\bibnamefont {Ding}}, \bibinfo {author} {\bibfnamefont {J.~M.}\ \bibnamefont {Morgado}}, \bibinfo {author} {\bibfnamefont {K.}~\bibnamefont {Raya}}, \bibinfo {author} {\bibfnamefont {D.}~\bibnamefont {Binosi}}, \bibinfo {author} {\bibfnamefont {L.}~\bibnamefont {Chang}}, \bibinfo {author} {\bibfnamefont {F.}~\bibnamefont {De~Soto}}, \bibinfo {author} {\bibfnamefont {C.~D.}\ \bibnamefont {Roberts}}, \bibinfo {author} {\bibfnamefont {J.}~\bibnamefont {Rodr\'\i{}guez-Quintero}}, \ and\ \bibinfo {author} {\bibfnamefont {S.~M.}\ \bibnamefont {Schmidt}},\ }\href {\doibase 10.1103/PhysRevD.105.L091502} {\bibfield  {journal} {\bibinfo  {journal} {Phys. Rev. D}\ }\textbf {\bibinfo {volume} {105}},\ \bibinfo {pages} {L091502} (\bibinfo {year} {2022})},\ \Eprint {http://arxiv.org/abs/2201.00884} {arXiv:2201.00884 [hep-ph]} \BibitemShut {NoStop}%
\bibitem [{\citenamefont {Raya}\ \emph {et~al.}(2022)\citenamefont {Raya}, \citenamefont {Cui}, \citenamefont {Chang}, \citenamefont {Morgado}, \citenamefont {Roberts},\ and\ \citenamefont {Rodriguez-Quintero}}]{Raya:2021zrz}%
  \BibitemOpen
  \bibfield  {author} {\bibinfo {author} {\bibfnamefont {K.}~\bibnamefont {Raya}}, \bibinfo {author} {\bibfnamefont {Z.-F.}\ \bibnamefont {Cui}}, \bibinfo {author} {\bibfnamefont {L.}~\bibnamefont {Chang}}, \bibinfo {author} {\bibfnamefont {J.-M.}\ \bibnamefont {Morgado}}, \bibinfo {author} {\bibfnamefont {C.~D.}\ \bibnamefont {Roberts}}, \ and\ \bibinfo {author} {\bibfnamefont {J.}~\bibnamefont {Rodriguez-Quintero}},\ }\href {\doibase 10.1088/1674-1137/ac3071} {\bibfield  {journal} {\bibinfo  {journal} {Chin. Phys. C}\ }\textbf {\bibinfo {volume} {46}},\ \bibinfo {pages} {013105} (\bibinfo {year} {2022})},\ \Eprint {http://arxiv.org/abs/2109.11686} {arXiv:2109.11686 [hep-ph]} \BibitemShut {NoStop}%
\bibitem [{\citenamefont {Yin}\ \emph {et~al.}(2023)\citenamefont {Yin}, \citenamefont {Xu}, \citenamefont {Cui}, \citenamefont {Roberts},\ and\ \citenamefont {Rodr\'\i{}guez-Quintero}}]{Yin:2023dbw}%
  \BibitemOpen
  \bibfield  {author} {\bibinfo {author} {\bibfnamefont {P.-L.}\ \bibnamefont {Yin}}, \bibinfo {author} {\bibfnamefont {Y.-Z.}\ \bibnamefont {Xu}}, \bibinfo {author} {\bibfnamefont {Z.-F.}\ \bibnamefont {Cui}}, \bibinfo {author} {\bibfnamefont {C.~D.}\ \bibnamefont {Roberts}}, \ and\ \bibinfo {author} {\bibfnamefont {J.}~\bibnamefont {Rodr\'\i{}guez-Quintero}},\ }\href {\doibase 10.1088/0256-307X/40/9/091201} {\bibfield  {journal} {\bibinfo  {journal} {Chin. Phys. Lett.}\ }\textbf {\bibinfo {volume} {40}},\ \bibinfo {pages} {091201} (\bibinfo {year} {2023})},\ \Eprint {http://arxiv.org/abs/2306.03274} {arXiv:2306.03274 [hep-ph]} \BibitemShut {NoStop}%
\end{thebibliography}%

\appendix

\section{Appendix A: Proofs of the sufficiency of strengthened sieves}

This appendix will show proofs of the sufficiency of the three types of strengthened sieves. These proofs will directly use the notations and theorems in Ref.~\cite{schmudgen2017moment} and also involve some functional analysis.

First, for the case of continuity, according to the Theorem 10.8 in Ref.~\cite{schmudgen2017moment}, what we need to prove is that each interior point of $\mathcal{S}_{m+1}$ has a continuous representing function. For the convenience of the following, we will prove this proposition on the general integration interval $[a,b]$. Consider a map $T$ from all continuous functions on the interval $[a,b]$ to $m+1$-dimensional real space:
\begin{equation}
T(f)=\left(\int_a^bf(x)\,dx,\int_a^bxf(x)\,dx,\cdots,\int_a^bx^mf(x)\,dx\right)\,,
\end{equation}
which is obviously a continuous linear map, and by the open mapping theorem is also an open map.

Let the set of all non-negative continuous functions on the interval $[a,b]$, except for the zero function, be $X$, and the set of all interior points of $\mathcal{S}_{m+1}$ be $Y$. Obviously, both $X$ and $T(X)$ are open convex sets, and $T(X)\subseteq Y$. Let the complement of $T(X)$ in $Y$ be $Z$, and we assume that $Z$ is not empty.
By Theorem 1.26 in Ref.~\cite{schmudgen2017moment}, each element in $Z$ has a representing function composed of several delta functions. The delta function can be the limit of a series of elements in $X$, so each element in $Z$ can be the limit of a series of elements in $T(X)$. Therefore, there must be an element of $T (X)$ in any neighborhood of the element in $Z$; that is, $Z$ has no interior points.
Consider a sufficiently small open ball $O$ centered on an element of $Z$, satisfying the condition $O\subseteq Y$. Obviously, $W=T(X)\bigcap O$ is a non-empty open set, so the set $W'$ that is symmetric about the center of $O$ with $W$ is also a non-empty open set. Since $T(X)$ is a convex set, $W'$ must belong to $Z$; otherwise, the center of $O$ would belong to $T(X)$. So $Z$ has a non-empty open subset, i.e., $Z$ has an interior point, which contradicts the previous conclusion. So $Z$ is an empty set, i.e., $T(X)=Y$, and the proof is complete.

Secondly, for the case of unimodality, according to Eqs.~\eqref{el1}, \eqref{el2}, Theorem 10.8, and Corollary 10.13 in Ref.~\cite{schmudgen2017moment}, there exists a series of canonical representing functions of $(\lambda t_k-t_{k+1})_{k=0}^m$. By appropriately selecting some canonical representing functions and combining them linearly, we can always obtain a representing function $h(x)$ that satisfies the following conditions: $h(\lambda)=0$ and the number of delta functions on both sides of $\lambda$ is greater than $m+1$.
Assume $h(x)=\sum_{j=1}^rm_j\delta(x-c_j)$. Consider the following function:
\begin{equation}
    w(x)=\sum_{j=1}^r\frac{m_j}{\lambda-c_j}\delta(x-c_j)-\left(\sum_{j=1}^r\frac{m_j}{\lambda-c_j}\right)\delta(x-\lambda)\,,
\end{equation}
and divide it into two parts: let the left part of $\lambda$ be $w^{+}(x)$, and the right part be $-w^{-}(x)$. Regarding the delta function at $\lambda$, if it is positive, it is included in $w^{+}(x)$; otherwise, it is included in $-w^{-}(x)$. Let the $k$-th moment of $w^{+}(x)$ be $t_k^{+}$, and the $k$-th moment of $w^{-}(x)$ be $t_k^{-}$. Obviously, there are $t_k^{+}-t_k^{-}=t_k$.
From the conditions satisfied by $h(x)$, we have: $ind(w^{+})>m+1$ and $ind(w^{+})>m+1$. Thus, by Theorem 10.7 in Ref.~\cite{schmudgen2017moment}, $(t_k^{+})_{k=0}^{m+1}$ and $(t_k^{-})_{k=0}^{m+1}$ are the interior points of $\mathcal{S}_{m+2}$ on $[0,\lambda]$ and $[\lambda,1]$, respectively. According to the first part of this appendix, both sequences have a continuous representing function, and let them be $g^{+}(x)$ and $g^{-}(x)$ respectively.  Combining $g^{+}(x)$ and $-g^{-}(x)$ and integrating them, we can get an unimodal representing function of $(s_k)_{k=0}^m$, and the peak location is $\lambda$.

Finally, for the case of symmetry, according to the previous proof, $(s_k)_{k=0}^m$ has an unimodal representing function, with the peak location $1/2$. Let such a representing function be $f(x)$ and consider the central moment of $f(x)$, which is related to the moments as follows:
\begin{equation}
    \int_0^1(x-1/2)^kf(x)\,dx=\sum_{i=0}^{k}(-1/2)^{k-i}\binom{k}{i}s_i\,.
\end{equation}
We see that moments can be converted into central moments, and in fact, the opposite process can be carried out similarly.
According to Eq.~\eqref{kl}, the odd-order central moments of $f(x)$ are zero up to $m$-th, so the first $m+1$ central moments of $f(1-x)$ are the same as that of $f(x)$. Since the central moment and moment can be converted to each other, $f(1-x)$ is also an unimodal representing function of $(s_k)_{k=0}^m$. Therefore, $[f(x)+f(1-x)]/2$ is a symmetric representation function of $(s_k)_{k=0}^m$.

\section{Appendix B: Probability distribution for moments of the symmetric PDF}

For the case of symmetric PDFs, their moments will obey different probability distributions due to Eq.~\eqref{kl}. Here, we take the case of $m=3$ as an example to illustrate. Using Eq.~\eqref{kl} and the PDF normalization, we get:
\begin{equation}
    s_0=1\,;\ \quad s_1=\frac{1}{2}\,;\ \quad s_3=\frac{3}{2}s_2-\frac{1}{4}\,.
\end{equation}
Four moments, three restrictions, thus leaving only one moment that can be considered as a random variable. Here we choose $s_2$ as the random variable and $s_3$ as the quantity derived from $s_2$. Thus, the probability distribution that $s_2$ obeys will be a mixture of $N(\mu_2,\sigma_2^2)$ and $N(\mu_3,\sigma_3^2)$. Specifically, in the asymmetric case, the probability distribution is $P(s_2,s_3)=N(\mu_2,\sigma_2^2)N(\mu_3,\sigma_3^2)$, and leaving aside the constant factor, it is:
\begin{equation}
    P(s_2,s_3)\propto \exp\left[-\frac{(s_2-\mu_2)^2}{2\sigma_2^2}-\frac{(s_3-\mu_3)^2}{2\sigma_3^2}\right]\,.
\end{equation}
Substituting the relationship between $s_2$ and $s_3$ into the above equation yields:
\begin{equation}
    P(s_2)\propto \exp\left[-\frac{(s_2-\mu_{2*})^2}{2\sigma_{2*}^2}\right]\,,
\end{equation}
where
\begin{subequations}
\begin{align}
\mu_{2*}=&\frac{8\mu_2\sigma_3^2+12\mu_3\sigma_2^2+3\sigma_3^2}{18\sigma_2^2+8\sigma_3^2}\,,\\
\sigma_{2*}=&\frac{2\sigma_2\sigma_3}{\sqrt{9\sigma_2^2+4\sigma_3^2}}\,.
\end{align}
\end{subequations}
Thus, $s_2$ still obeys a normal distribution with slightly more complex parameters. In general, if more moments are taken into account, under the restriction of Eq.~\eqref{kl}, the independent normal distributions followed by the moments will mix into a more complex multivariate normal distribution. 

\section{Appendix C: All-orders evolution of PDF moments and error propagation}
Assuming the existence of universality of QCD effective charge~\cite{Raya:2021zrz,Yin:2023dbw}, the evolution of the non-singlet PDF moments from an arbitrary scale $\zeta$ to hadron scale $\zeta_H$ can be completely determined by the value of the first moment of pion, namely,
\begin{equation}
    \langle x^m\rangle_M^{\zeta_{\mathcal{H}}}=\frac{\langle x^m\rangle_M^{\zeta}}{(2\langle x\rangle_{\pi}^{\zeta})^{\gamma_0^m/\gamma_0^1}}\,,
\end{equation}
where $\langle x^m\rangle_M^{\zeta}$ represents the $m$-th moment of the light quark valence PDF in meson $M$ (pion or kaon) at scale $\zeta$, and
\begin{equation}
    \gamma_0^m=-\frac{4}{3}\left[3+\frac{2}{(m+1)(m+2)}-4\sum_{j=1}^{m+1}\frac{1}{j}\right]\,.
\end{equation}
Regarding the evolution of the moment error, we compute the error at hadron scale according to the standard error propagation formula. Specifically, when $m=1$ and $M=\pi$, the error at scale $\zeta_{\mathcal{H}}$ is exactly zero; the errors in other cases are as follows:
\begin{equation}
    \sigma_{m,M}^{\zeta_{\mathcal{H}}}=\frac{\sqrt{(\sigma_{m,M}^{\zeta})^2+\left(-\frac{\gamma_0^m\langle x^m\rangle_{M}^{\zeta}}{\gamma_0^1\langle x\rangle_{\pi}^{\zeta}}\sigma_{1,\pi}^{\zeta}\right)^2}}{(2\langle x\rangle_{\pi}^{\zeta})^{\gamma_0^m/\gamma_0^1}}\,,
\end{equation}
where $\sigma_{m,M}^{\zeta}$ is the error of $\langle x^m\rangle_M^{\zeta}$. 
It is worth noting that for maximum consistency, the $\langle x\rangle_{\pi}^{\zeta}$ and $\sigma_{1,\pi}^{\zeta}$ required in the error propagation formula for a given dataset are chosen from the same source, i.e., we use $0.2541(26)$ for $\mathcal{A}$ and $0.261(3)$ for $\mathcal{B}$ and $\mathcal{C}$.

\end{document}